\documentclass[aps,prl,twocolumn,groupedaddress,showpacs]{revtex4}

\usepackage{graphicx}

\begin{document}

\title{$N$-particle scattering matrix for electrons
interacting on a quantum dot}

\author{A.V.\ Lebedev$^{\,a,b}$, G.B.\ Lesovik$^{\,b}$, and G.\ Blatter$^{\,a}$}

\affiliation{$^{a}$Theoretische Physik, Schafmattstrasse 32,
ETH-Zurich, CH-8093 Z\"urich, Switzerland}

\affiliation{$^{b}$L.D.\ Landau Institute for Theoretical Physics,
RAS, 119334 Moscow, Russia}

\date{\today}

\begin{abstract}

We present a non-perturbative expression for the scattering matrix of $N$
particles interacting inside a quantum dot. Characterizing the dot by its
resonances, we find a compact form for the scattering matrix in a real-time
representation.  We study the transmission probabilities and
interaction-induced orbital entanglement of two electrons incident on the dot
in a spin-singlet state.

\end{abstract}


\maketitle

The scattering matrix, taking asymptotically free incoming states through an
interaction region and providing the free outgoing states, is of huge basic
and practical interest.  Originally introduced by Born \cite{born} and by
Wheeler and independently by Heisenberg \cite{wheeler} in atomic and particle
physics, its application to electron transport \cite{landauer} has made it
into a central tool of mesoscopic physics \cite{imry99}.  Its formulation for
non-interacting electrons provides the two-terminal conductance between
reservoirs in terms of the transmission probability across the scatterer
\cite{landauer} and generalizations to complex setups \cite{buettikermt} have
been successful. Even more, the scattering matrix approach has been used to
study noise \cite{noise} and a description of fluctuations in non-equilibrium
situations is provided by the full counting statistics \cite{levitov93}, with
numerous applications known today \cite{review}.  These successes ask for a
generalization to the interacting case \cite{imry}, for which Goorden {\it et
al.} \cite{goorden07} have recently derived the scattering matrix of two
coupled conductors within a perturbative calculation.

Of generic interest in mesoscopic physics is the quantum dot, where the
transport across is usually studied for a finite bias, and different
approximate schemes have been used to include interactions in the transport
analysis \cite{FCSint}. Recent interest concentrates on a new type of
transport, where designed voltage pulses are generated to send a finite number
of electrons towards the scattering region \cite{pulses}, see Ref.\
\cite{glattli} for recent experiments in this direction.  In this situation,
novel effects, such as those due to particle exchange, show up quite
prominently \cite{hassler}.  Moreover, this type of control on individual
particles allows to study fundamental quantum properties such as the
entanglement between electrons \cite{lebedev}.

In this letter, we derive the $N$-particle scattering matrix for electrons
that propagate freely in the leads but exhibit Coulomb repulsion when
interacting on the dot. Within our formalism, the interaction is accounted for
by the Hamiltonian $\hat H_\mathrm{int} = e^2\hat N^2/2C$, where $\hat N$ is
the dot's electron number operator and $C$ denotes its capacitance.  We use
our results to study the wave function and degree of entanglement of two
scattered electrons.  Below, we construct the two-particle scattering matrix
for the quantum dot including Coulomb interaction and generalize the result to
the $N$-particle situation.

Usually, the scattering matrix connects states at given energies; here, we
start with the propagator describing the scattering of wave packets in
coordinate space. We start with a (properly symmetrized, spin indices are
suppressed) incident two-electron wave function at time $t_1$,
$\Psi^\mathrm{in}(\vec y,t_1)$, with $\vec y=\{y_1,y_2\}$. The scattered wave
function at later times $t_2 > t_1$ can be obtained with the help of the
two-particle propagator $K^{\scriptscriptstyle (2)}(\vec x,t_2;\vec y,t_1)$
describing the evolution of two particles from the initial positions $\vec y$
at time $t_1$ to the final positions $\vec x$ at $t_2$,
\begin{equation}
      \Psi^\mathrm{out}(\vec x,t_2)
      =\int d^2y \,  K^{\scriptscriptstyle (2)}
      (\vec x,t_2; \vec y,t_1)\,
      \Psi^\mathrm{in}(\vec y,t_1).
      \label{out}
\end{equation}
The two-particle propagator $K^{\scriptscriptstyle (2)}$ can be defined
through a Feynman path integral over trajectories $\vec x(t)$,
\begin{equation}
      K^{\scriptscriptstyle(2)}(\vec{x},t_2;\vec{y},t_1) \!=\!\!
      \int\! {\cal D}[\vec{x}\,]\,
      \exp \Bigl( \frac{i}{\hbar} \int_{t_1}^{t_2}\!\!dt\,
      L^{\scriptscriptstyle (2)}(\vec{x};\dot{\vec{x}}\,) \Bigr),
\end{equation}
with the boundary conditions $\vec x(t_1) = \vec y$. Here,
$L^{\scriptscriptstyle (2)}(\vec{x}, \dot{\vec{x}}\,)$ is the system's
Lagrangian including kinetic ($\propto m$), dot potential ($U$), and
interaction ($\propto U_c = 2e^2/C$) energies,
\begin{equation}
      L^{\scriptscriptstyle (2)} = \!
      \sum_{i=1}^2 \Bigl[\frac{m \dot{x}_i^2}{2} - U(x_i) \Bigr]
      \!-\!  \frac{U_c}{4}
      \bigl[\chi_\mathrm{d}(x_1)\!+\!\chi_\mathrm{d}(x_2) \bigr]^2,
      \label{lag}
\end{equation}
with the characteristic function $\chi_\mathrm{d}(x)$ of the dot equal to
unity within the dot and zero outside \cite{cap_form}.  

Without interaction, the two-particle propagator factorizes,
$K^{\scriptscriptstyle (2)}(\vec x,t_2;\vec y,t_1) = \Pi_i
K^{\scriptscriptstyle (1)}(x_i,t_2;y_i,t_1)$ with $K^{\scriptscriptstyle
(1)}(x,t_2;y,t_1)$ the one-particle propagator, while the interaction mixes
the particle trajectories.  A Hubbard-Stratonovich transformation \cite{HS}
with the real auxiliary field $z(t)$ allows us to decouple the quadratic
interaction
\begin{eqnarray}
      &&K^{\scriptscriptstyle (2)}(\vec{x},t_2;\vec{y},t_1)
      = \int {\cal D}[z] \exp \Bigl[i\frac{U_c}{\hbar}
      \int dt\, z^2(t) \Bigr] \label{zav}\\
      &&\qquad\times\,
      K_{[z]}^{\scriptscriptstyle (1)}(x_1,t_2;y_1,t_1) \,
      K_{[z]}^{\scriptscriptstyle (1)}(x_2,t_2;y_2,t_1),
      \nonumber
\end{eqnarray}
where $K_{[z]}^{\scriptscriptstyle (1)}(x,t_2;y,t_1)$ is the one-particle
propagator in the presence of a fluctuating potential $U_c(t) = U_c z(t)$,
\[
      K^{\scriptscriptstyle (1)}_{[z]}
      =\int {\cal
      D}[x] \exp\Bigl[ \frac{i}{\hbar} \int_{t_1}^{t_2} \!\!\! dt \,
      \Bigl(\frac{m\dot{x}^2}{2}\!-U(x)\!-U_c z(t)\chi_\mathrm{d}(x)\Bigr)
      \Bigr].
\]

Next, we introduce the scattering matrix $S_{\alpha\beta}^{\scriptscriptstyle
(1)}(\varepsilon)$ of the dot in the absence of the fluctuating potential
$U_c(t)$; the indices $\alpha,\beta \in\{{\scriptstyle\rm L,R}\}$ specify the
lead indices for the outgoing ($\alpha$) and incoming ($\beta$) scattering
channels and $\varepsilon$ denotes the energy variable. We describe the dot
through the resonance positions ($\epsilon_j$) and (identical) widths
($\Gamma$); the scattering matrix $S_{\alpha\beta}^{\scriptscriptstyle (1)}
(\varepsilon)$ then takes the form
\begin{equation}
      S_{\alpha\beta}^{\scriptscriptstyle (1)}(\varepsilon) =
      r_{\alpha\beta} +
      \sum_{j} \frac{i\Gamma/2}{\varepsilon-\epsilon_j +
      i\Gamma/2}\, s_{\alpha\beta}^{\scriptscriptstyle (j)},
      \label{1sc_en}
\end{equation}
where the constant $2\times 2$ matrices $r_{\alpha\beta}$ and $s_{\alpha
\beta}^{\scriptscriptstyle (j)}$ can be found from the unitarity conditions.
The Fourier transform provides the real time ($\tau$) representation
\begin{equation}
      S_{\alpha\beta}^{\scriptscriptstyle (1)}(\tau)
      = \delta(\tau) r_{\alpha\beta}
      + \theta(\tau) \sum_j\frac{\eta}{2} e^{-i\omega_j\tau}
      e^{-\eta\tau/2} \, s_{\alpha\beta}^{\scriptscriptstyle (j)},
      \label{1sc_time}
\end{equation}
where $\eta = \Gamma/\hbar$ is the inverse dwell time, $\omega_j=
\epsilon_j/\hbar$ is the resonance frequency, and $\delta(\tau)$,
$\theta(\tau)$ are the usual $\delta$- and Heaviside functions.  The first
term in Eq.~(\ref{1sc_time}) describes the reflection of a particle that has
not penetrated into the dot, while the subsequent terms correspond to
processes where the particle has spent a time $\tau$ inside the dot; the
factor $e^{-i\omega_j\tau}$ describes the accumulated phase. The presence of
the fluctuating potential $U_c(t)$ contributes an additional phase to the
one-particle scattering matrix (\ref{1sc_time}),
\begin{equation}
      S_{\alpha\beta,[z]}^{\scriptscriptstyle (1)}(t_2,t_1) =
      S_{\alpha\beta}^{\scriptscriptstyle (1)}(t_2-t_1)
      \exp \Bigl( - \frac{i}{\hbar} \int_{t_1}^{t_2} U_c(t) dt
      \Bigr),
      \label{1sc_time_z}
\end{equation}
where $t_1$ and $t_2$ denote the arrival and escape times of the particle (we
assume escape amplitudes that depend weakly on energy).  The additional phase
derives from the gauge transformation $\Psi_z(x,t) \rightarrow \Psi(x,t)
\exp(-\frac{i}{\hbar} \int^{t} dt'\, U_c(t'))$; we neglect the scattering
potential of size $\sqrt{U_c \, \hbar v/l_\mathrm{d}} \sim U_c \sqrt{v
\epsilon/\alpha c}$ arising at the dot's edge (see Ref.\ \onlinecite{time};
here, $l_\mathrm{d}$ and $\epsilon$ denote the dot size and its dielectric
constant, $\alpha$ is the fine structure constant).  With a typical mesoscopic
setup in mind, we assume velocities $v$ of order of the Fermi velocity and
energies larger than the Coulomb energy, $\varepsilon(k)\gg U_c$; with typical
ratios $v/c \sim 10^{-2}$, we can safely ignore the fluctuation corrections in
the dot potential $U(x)$.

Next, we express the propagator $K^{\scriptscriptstyle (1)}_{[z]}$ through the
scattering matrix~(\ref{1sc_time_z}). To simplify matters, we linearize the
spectrum, $\varepsilon(k) = \hbar v k$; a particle escaped out of the dot then
never returns.  In terms of trajectories, the scattering process involves
three stages: {\it i)} the ballistic motion with velocity $v$ towards the dot,
{\it ii)} the dwell time in the dot, and, {\it iii)} the ballistic propagation
away from the dot.  We define the coordinates in the left ($x<0$) and right
($x>0$) leads with respect to the left ($x=0^-$) and right ($x=0^+$) dot
boundaries and express the propagator $K_{[z]}^{\scriptscriptstyle (1)}$
through the scattering matrix~(\ref{1sc_time_z}), $K^{\scriptscriptstyle
(1)}_{\alpha\beta,[z]} (x,t_2;y,t_1) = S_{\alpha\beta,[z]}^{\scriptscriptstyle
(1)}(\tau,s)/v$, where $s = t_1+|y|/v$ and $\tau = t_2-|x|/ v$ are the arrival
and escape times of the particle to and from the dot; similar definitions
($s_i=t_1+|y_i|/v$ and $\tau_i = t_2-|x_i|/v$) apply to the two-particle
scattering matrix, for which we write
\begin{widetext}
\begin{equation} 
      S^{\scriptscriptstyle (2)}_{\alpha_1\alpha_2
      \beta_1\beta_2}(\tau_1,\tau_2;s_1,s_2) =
      S^{\scriptscriptstyle(1)}_{\alpha_1\beta_1}(\tau_1\!-\!s_1)
      S^{\scriptscriptstyle(1)}_{\alpha_2\beta_2}(\tau_2\!-\!s_2)
      \Bigl\langle \exp\Bigl\{ -i\omega_c
      \Bigl[\int_{s_1}^{\tau_1} dt\, z(t)+
             \int_{s_2}^{\tau_2} dt\, z(t) \Bigr]
      \Bigr\}\Bigr\rangle,
      \label{1pr_z3}
\end{equation}
and in terms of which the two-particle propagator (\ref{zav}) assumes the form
$K^{\scriptscriptstyle(2)}= S^{\scriptscriptstyle (2)}/v^2$.  Here, $\omega_c
= U_c/\hbar$ and the average in Eq.\ (\ref{1pr_z3}) is taken with respect to
the fluctuating Gaussian field $z(t)$.  The latter is $\delta$ correlated in
time, $\langle z(t_2) z(t_1) \rangle = (i/2\omega_c) \delta(t_2-t_1)$ (the
complex propagator in Eq.~(\ref{zav}) generates an imaginary correlator for
the real field $z(t)$) and thus the last factor in Eq.~(\ref{1pr_z3}) can be
explicitly averaged over with the result
\begin{equation} \label{2sc_time}
      S^{\scriptscriptstyle (2)}_{\alpha_1\alpha_2
      \beta_1\beta_2}(\tau_1,\tau_2;s_1,s_2)
      = \tilde S^{\scriptscriptstyle(1)}_{\alpha_1\beta_1}
      (\tau_1\!-\!s_1)
      \tilde S^{\scriptscriptstyle(1)}_{\alpha_2\beta_2}
      (\tau_2\!-\!s_2)
      \exp (-i\omega_c\tau_{12}/2),
\end{equation}
\end{widetext}
where $\tilde S^{(1)}_{\alpha\beta}$ is the scattering matrix~(\ref{1sc_time})
with renormalized resonance frequencies $\tilde\epsilon_j =\epsilon_j +
U_c/4$ and $\tau_{12}$ is the time the two particles spend together in
the dot,
\[
      \tau_{12} = {\scriptstyle \frac12} (|\tau_1-s_2|+|\tau_2-s_1| -
      |\tau_1-\tau_2| - |s_1-s_2|).
\]
This two-particle scattering matrix (\ref{2sc_time}) is the key result of
this Letter. All effects of Coulomb interaction are accounted for by
renormalized resonance energies due to self-interaction of individual
electrons in the dot and an additional phase accumulated by the electrons
during their simultaneous presence in the quantum dot. 

An inverse Fourier transformation provides us with the energy representation
\begin{widetext}
\begin{eqnarray} \label{2sc_energy}
      &&S^{\scriptscriptstyle (2)}_{\alpha_1\alpha_2
      \beta_1\beta_2}(\varepsilon_1^\prime,\varepsilon_2^\prime;
      \varepsilon_1,\varepsilon_2)
      =(2\pi)^2
      \delta(\varepsilon_1\!-\!\varepsilon_1^\prime)
      \delta(\varepsilon_2\!-\!\varepsilon_2^\prime)
      \, S_{\alpha_1\beta_1}^{\scriptscriptstyle (1)}(\varepsilon_1)
      S_{\alpha_2\beta_2}^{\scriptscriptstyle (1)}(\varepsilon_2)
      + 2\pi\delta(\varepsilon_1\! +\! \varepsilon_2\! -\!
      \varepsilon_1^\prime\! -\! \varepsilon_2^\prime)\\
      \nonumber
      && \qquad\times \sum_{jk}
      \, \frac{(iU_c/2)\,
      s_{\alpha_1\beta_1}^{\scriptscriptstyle(j)}
      s_{\alpha_2\beta_2}^{\scriptscriptstyle(k)}}{
      \varepsilon_1\! +\! \varepsilon_2\! -\!
      \tilde\epsilon_j\!-\!\tilde\epsilon_k\!-\! U_c/2\!+\!i\Gamma}
      \frac{i\frac{\Gamma}{2}}{\varepsilon_1\! -\! \tilde\epsilon_j
      \!+\!i\frac{\Gamma}{2}}
      \frac{i\frac{\Gamma}{2}}{\varepsilon_2\! -\! \tilde\epsilon_k
      \!+\!i\frac{\Gamma}{2}}
      \biggl( \frac{1}{\varepsilon_1^\prime\! -\! \tilde\epsilon_j
      \!+\!i\frac{\Gamma}{2}}
      + \frac{1}{\varepsilon_2^\prime\! -\! \tilde\epsilon_k
      \!+\!i\frac{\Gamma}{2}}
      \biggr).
\end{eqnarray}
\end{widetext}
The first term describes the non-interacting process where particles scatter
sequentially. The second term accounts for inelastic processes where only the
total energy $E =\varepsilon_1+\varepsilon_2$ is conserved. The Coulomb
interaction generates additional poles at $E_{jk}=\tilde\epsilon_j
+\tilde\epsilon_k +U_c/2-i\Gamma$ involving the total energy $E$.
These interaction-induced singularities cannot be obtained via a
perturbative expansion for large $U_c\gg\Gamma$. For weak interaction $U_c\ll
\Gamma$ or far away from the resonances $|E-\tilde\epsilon_j -
\tilde\epsilon_k|\gg U_c$, the expansion of Eq.~(\ref{2sc_energy}) to first
order in $U_c$ reproduces the perturbative result obtained in
Ref.~\cite{goorden07}.

The above derivation for the two-particle scattering matrix can be generalized
to $N$ particles; the averaging over the field $z(t)$ generates an
additional phase factor accounting for the pairwise interaction of particles
residing simultaneously (for a time $\tau_{jk}$) on the dot,
\[
      S_{\{\alpha_j\beta_j\}}^{\scriptscriptstyle(N)}
      \bigl(\{\tau_j;s_j\}\bigr) =
      \prod_{j> k}^N
      e^{-i\omega_c \tau_{jk}/2} \prod_{j=1}^N
      \tilde S^{\scriptscriptstyle (1)}_{\alpha_j
      \beta_j}(\tau_j\!-\!s_j).
\]
The above result also holds true for a multichannel setup, with
$\alpha_j,~\beta_j$, $j=1,\dots,N$ turning into multichannel indices. In
particular, the results can be straightforwardly applied to the experimental
setup \cite{marcus07} with two parallel leads feeding/emptying two
capacitively coupled dots that has been recently used to measure
interaction-induced cross correlations, see also Ref.~\onlinecite{goorden07}.

In applying our results to realistic mesoscopic problems, we have to avoid
mixing between the scattered particles and the electrons in the Fermi sea.
Hence, we do not consider situations with levels within the distance $\Gamma$
around the Fermi energy $\varepsilon_{\rm \scriptscriptstyle F}$ and assume
that $U_c$ does not shift a level across $\varepsilon_{\rm \scriptscriptstyle
F}$; the latter allows us to ignore complications due to the Kondo effect
\cite{GM}.  In the following, we study the scattering problem of two
single-electron excitations created above the Fermi sea and a quantum dot with
only one resonance at $\tilde\epsilon_0$ above the Fermi energy
$\varepsilon_{\rm \scriptscriptstyle F}$, $\tilde\epsilon_0-\varepsilon_{\rm
\scriptscriptstyle F} \gg \Gamma$. The scattering matrix (\ref{2sc_energy})
then tells, that (the non-trivial component of) the scattered wave function
involves energies near $\tilde\epsilon_0$ and $\tilde\epsilon_+ =
\tilde\epsilon_0 + U_c/2$.

We start from a two-electron state with wave function $\Psi^\mathrm{in}
(x_1,x_2)$ created at time $t=0$ in the left lead and moving towards to the
dot \cite{Fermisea}. The scattered wave is given by Eq.~(\ref{out}) and can be
expressed in terms of retarded variables $\xi_{1,2} = |x_{1,2}| - v_{\rm
\scriptscriptstyle F} t$, with $v_{\rm \scriptscriptstyle F}$ the Fermi
velocity.  The scattered wave to the right of the dot involving tunneling of
both electrons assumes the form ($Y\equiv y1+y_2$)
\begin{widetext}
\begin{eqnarray}\label{2waveRR}
      &&\Psi_{\scriptscriptstyle\rm RR}(\xi_1,\xi_2) =
      \frac{s_{\scriptscriptstyle\rm RL}^2}{\ell^2}
      \biggl[ e^{ik_+\xi_>} e^{ik_0 \xi_<}
      \int_{\xi_>}^0 \!\!\! dy_1 dy_2\,
      \Psi^\mathrm{in}(y_1,y_2)
      e^{-ik_c(Y-|y_1-y_2|)/2} e^{-ik_0 Y}
      e^{(\xi_1+\xi_2-Y)/\ell}
      \\ 
      &&\qquad\qquad  + e^{ik_0(\xi_1+\xi_2)} 
      \!\int_{\xi_<}^{\xi_>}\!\!\!\! dy_1
      \!\int_{\xi_>}^0 \!\!\! dy_2\, 
      e^{-ik_0 Y} e^{(\xi_1+\xi_2-Y)/\ell}
      \bigl[
      \theta(\xi_2\!-\!\xi_1)\Psi^\mathrm{in}(y_1,y_2)
      +\theta(\xi_1\!-\!\xi_2) \Psi^\mathrm{in}(y_2,y_1)
      \bigr] \biggr], \nonumber
\end{eqnarray}
\end{widetext}
where $\ell=2\hbar/\Gamma v_{\rm\scriptscriptstyle F}$ is the real-space width
of the scattered wave, $\xi_>=\max\{\xi_1,\xi_2\}$, $\xi_<=\min\{\xi_1,
\xi_2\}$, $k_0 = \tilde\omega_0/v_{\rm\scriptscriptstyle F}$, $k_c = \omega_c/
2v_{\rm \scriptscriptstyle F}$, and $k_+ = k_0+k_c$. The second term describes
the process where the electrons do not overlap in the dot, while the term
$\propto e^{ik_+\xi_>} e^{ik_0\xi_<}$ deals with the case where both electrons
occupy the dot simultaneously during scattering.  For electrons in a
spin-triplet state with anti-symmetric orbital wave function $\Psi^\mathrm{in}
(y_1,y_2)$, this term vanishes and no interaction effects survive, a
consequence of the Pauli principle.

Next, we choose a spin-singlet incoming state with one orbit $\phi$, hence
$\Psi^\mathrm{in}(y_1,y_2) = \phi(y_1) \phi(y_2)$. To simplify matters, we
choose an exponentially truncated plane wave function with a wave vector in
resonance with the dot, $\phi(x) = \theta(a-x)\, e^{ik_0(x-a)}
e^{(x-a)/2L}/\sqrt{L}$, where $a<0$ is the initial position of the wave packet
at $t=0$ and $L$ denotes its width; this choice allows to perform all
integrals in Eq.~(\ref{2waveRR}) explicitly.  We determine the probabilities
$P_1$ and $P_2$ to transmit one and two electrons through the dot (at
resonance, where $|s_{\scriptscriptstyle\rm RL}|=1$) and find them to depend
only on the two dimensionless parameters $\alpha = k_c L$ and $\beta =
\ell/2L$ quantifying the interaction and the resonance width, respectively,
\begin{eqnarray}
      &&P_1 =\frac{2\beta}{1+3\beta} \biggl[
      \frac{2+3\beta}{(1+\beta)^2}- \frac{1}{(\alpha\beta)^2+(1+\beta)^2}
      \biggr],
      \label{p1}\\
      &&P_2 = \frac{1}{(1\!+\!3\beta)(1+\beta)^2} \biggl[1+\beta(3+\beta)
      \label{p2}
      \\
      &&\quad \times\frac{4(\alpha\beta)^2 +
      3(\beta\!+\!3)(\beta\!+\!1)^2}{
      ((\alpha\beta)^2 +(1\!+\!\beta)^2)( 4(\alpha\beta)^2 +(3\!+\!\beta)^2)}
      \biggr].
      \nonumber
\end{eqnarray}
Without Coulomb interaction ($\alpha = 0$) or for vanishing dwell time
($\beta=0$), we recover the results $P_1=2p(1-p)$ and $P_2=p^2$, with
$p=1/(1+\beta)$ the single-particle tunnelling probability. Both, Coulomb
interaction ($\alpha >0$) and finite dwell time ($\beta >0$) suppress the
probability $P_2$ as compared with the non-interacting value $p^2$.  Even
infinite Coulomb energy ($\alpha \to \infty$) still permits tunnelling of two
electrons through the dot via sequential tunneling.

Finally, we show that the Coulomb interaction in the dot leads to an orbital
entanglement of the two particles (for interaction-induced spin entanglement
in a quantum dot, see Ref.\ \onlinecite{oliver}). Here, we concentrate on the
component of the wave function where two electrons are transmitted to the
right and estimate its degree of entanglement, which is entirely due to the
interaction in the dot.  We analyze the situation where the length $L$ of the
incoming wave packet tends to zero, hence $\beta \rightarrow \infty$. In this
case the normalized wave function on the right has the universal form
(independent of $\Psi^\mathrm{in}$)
\begin{equation}
      \Psi_{\scriptscriptstyle\rm RR}
      (\xi_1,\xi_2) =(2/\ell)\,
      e^{ik_+ \xi_>} e^{ik_0 \xi_<} \,
      e^{(\xi_1+\xi_2)/\ell},
      \label{2wave_sg}
\end{equation}
where $\xi_{1,2} = x_{1,2}\!-\!a\!-\!v_{\scriptscriptstyle\rm F}t<0$.
Eq.~(\ref{2wave_sg}) describes a two-electron state with different momenta
$k_+$ and $k_0 < k_+$, as has to be expected since the first electron escaping
carries an energy shifted up by the Coulomb interaction. The
state~(\ref{2wave_sg}) can be rewritten in a form
\begin{equation}
      \Psi_{\scriptscriptstyle\rm RR}
      =(2/\ell)
      e^{i(k_0+\frac{k_c}{2})(\xi_1+\xi_2)}\, e^{i\frac{k_c}{2}|\xi_1-\xi_2|}
      e^{(\xi_1+\xi_2)/\ell},
      \label{2wave_epr}
\end{equation}
reminding about the original Einstein-Podolsky-Rosen state with orbital
entanglement \cite{EPR}. To quantify its entanglement, one may calculate the
von Neumann entropy $E$ of the reduced density matrix $\rho(x,x^\prime) = \int
dx_2\, \Psi_{\scriptscriptstyle\rm RR}(x,x_2) \Psi_{\scriptscriptstyle\rm
RR}^*(x^\prime,x_2)$. Instead, we determine the purity $\Pi(\rho) =
\mbox{tr}\, \rho^2$, which is unity for separable states and provides the
lower limit $E > 1-\Pi$. With $A \equiv ik_c\ell/ (2-ik_c\ell)$, we find the
density matrix
\begin{eqnarray} \label{density}
      \rho(\xi,\xi^\prime) &=& (2/\ell)\,\theta(-\xi)
      \theta(-\xi^\prime) e^{(\xi+\xi^\prime)/\ell}
      e^{ik_0(\xi-\xi^\prime)}\\
      &&\>\times \bigl[ 1
      + A\,\theta(\xi\!-\!\xi^\prime) (e^{2\xi/\ell}-
      e^{ik_c(\xi-\xi^\prime)}  e^{2\xi^\prime/\ell})
      \nonumber\\
      &&\quad
      + A^*\theta(\xi^\prime\!-\!\xi) (e^{2\xi^\prime/\ell}-
      e^{ik_c(\xi-\xi^\prime)} e^{2\xi/\ell}) \bigr], \nonumber
\end{eqnarray}
that results into a purity $\Pi = [1+2/(1+(k_c\ell/4)^2)]/3$.  We conclude
that at finite $U_c$ the state~(\ref{2wave_epr}) is entangled and the degree
of entanglement saturates as the Coulomb interaction becomes larger than the
resonance width, $k_c\ell = U_c/\Gamma \gg 1$, i.e., when the energies of the
escaped particles become distinguishable.

In conclusion, our expression for the multi-particle scattering matrix
accounts for the redistribution of particle energies during inelastic
scattering with the appearance of new resonance poles that cannot be obtained
perturbatively for large $U_c \gg \Gamma$. As an application, we have studied
the case where two electrons are transmitted across a dot with a single
resonance and have investigated the ensuing orbital entanglement and the
reduction in the two-particle transmission due to the interaction.

We thank Fabian Hassler and Konstantin Matveev for discussions and acknowledge
the financial support from the Swiss National Foundation (through the program
MaNEP and the CTS-ETHZ), the Russian Foundation for Basic Research
(08-02-00767a), and the program 'Quantum Macrophysics' of the RAS.


\begin{thebibliography}{99}
\bibitem{cap_form} The capacitive form implies an instantaneous and uniform
change of the dot potential upon tunneling of the electron; the approximation
is valid for dots with plasma frequencies well above ballistic ones
(of order GHz).
\bibitem{Fermisea} Use of unit-flux voltage pulses guarantees that
excited particles are not entangled with the Fermi sea.
\end{thebibliography}
\end{document}